\documentclass{article}
\usepackage[T1]{fontenc}
\usepackage[utf8]{inputenc}
\usepackage{ismir} 
\usepackage{amsmath,cite,url}
\usepackage{graphicx}
\usepackage{color}

\usepackage{kotex}
\usepackage{enumitem}

\usepackage{multirow}
\usepackage{textcomp}
\usepackage{siunitx}
\usepackage{musicography}
\usepackage{booktabs} 
\usepackage{float} 

\title{On the de-duplication of the Lakh MIDI dataset}

\multauthor
  {Eunjin Choi$^1$ \hspace{1cm} Hyerin Kim$^2$ \hspace{1cm} Jiwoo Ryu$^2$}
  {{\bf Juhan Nam$^1$ \hspace{1cm} Dasaem Jeong$^2$}\\
  $^1$ Graduate School of Culture Technology, KAIST, South Korea\\
  $^2$  Department of Art \& Technology, Sogang University, South Korea\\
  {\tt\small \{jech,juhan.nam\}@kaist.ac.kr, \{kime0225, clayryu338\}@gmail.com, \{dasaemj\}@sogang.ac.kr}
  }

\def\authorname{F. Author, S. Author, and T. Author}

\begin{document}

\maketitle

\begin{abstract}
A large-scale dataset is essential for training a well-generalized deep-learning model. Most such datasets are collected via scraping from various internet sources, inevitably introducing duplicated data.
In the symbolic music domain, these duplicates often come from multiple user arrangements and metadata changes after simple editing. 
However, despite critical issues such as unreliable training evaluation from data leakage during random splitting, dataset duplication has not been extensively addressed in the MIR community. This study investigates the dataset duplication issues regarding Lakh MIDI Dataset (LMD), one of the largest publicly available sources in the symbolic music domain. To find and evaluate the best retrieval method for duplicated data, we employed the Clean MIDI subset of the LMD as a benchmark test set, in which different versions of the same songs are grouped together. We first evaluated rule-based approaches and previous symbolic music retrieval models for de-duplication and also investigated with a contrastive learning-based BERT model with various augmentations to find duplicate files. As a result, we propose three different versions of the filtered list of LMD, which filters out at least 38,134 samples in the most conservative settings among 178,561 files.
\end{abstract}

\section{Introduction}\label{sec:introduction}
As data-driven approaches become mainstream and huge generative neural networks become popular, the significance of large-scale datasets is increasing. The tremendous performance of generative models has been attributed to massive dataset availability. 
For example, the availability of billions of text-image pair data boosted the high-quality image synthesis from text in the computer vision domain~\cite{jia2021scaling}. Among several dataset collection strategies, web scraping~\cite{RaffelLMDphd}, or synthesis~\cite{manilow2019cutting_slakh} has emerged as the most practical method for assembling large-scale datasets, given the high costs associated with manual collection.

However, collecting large-scale datasets via web crawling can inevitably cause problems, including privacy, copyright, and data duplication issues. In this paper, we focus on data duplication, particularly highlighting issues in the use of the Lakh MIDI Dataset (LMD)~\cite{RaffelLMDphd}, which is widely recognized as one of the large-scale datasets in MIR. While there are studies in other fields that discuss the downsides of dataset duplication \cite{webster2023duplication} and the benefits of de-duplication \cite {lee2021deduplicatingbetter}, we found that such discussions are notably lacking in the music information retrieval (MIR) community. One related work is the examination of dataset validity issues in GTZAN \cite{sturm2013gtzan, sturm2014state}, and our work shares this intent by addressing the correct usage of LMD through de-duplication.

In particular, we argue that the current use of the LMD in symbolic music generation impairs the validity of existing experiments. The issue arises from unreliable training evaluations caused by data leakage during random splits. 
Duplicates across training, validation, and test splits can skew evaluation metrics such as cross-entropy loss, which many previous studies used to assess their model performance \cite{liang2019midi-sandwich2, donahue2019lakhnes, ren2020popmag, yu2022museformer, von2022figaro, thickstun2023anticipatory}. In the music generation domain, subjective evaluation is costly, and objective evaluation metrics are often insufficient to fully capture the quality of generated music. Consequently, studies rely on validation loss as a metric to claim non-overfitting and model effectiveness \cite{thickstun2023anticipatory}. 
Duplicates can also bias listening-based evaluations, particularly when the test split is used for conditioning. This is common practice in conditional music generation~\cite{ren2020popmag, von2022figaro, thickstun2023anticipatory, bhandariText2midiGeneratingSymbolic2025}.
If the duplication remains unaddressed, random splits that lead to data leakage will continue to undermine the reliability of evaluations. 

However, manually finding duplicates in a large-scale dataset such as LMD is virtually infeasible. Therefore, we explored cleaning LMD using rule-based and neural approaches. Our contributions are as follows: 

\begin{itemize}[leftmargin=10pt]
    \item \textbf{How to clean the LMD?} \\
We evaluated rule-based methods and existing symbolic music retrieval models for duplicate detection. We also explored training an unsupervised contrastive BERT model with augmentations to detect duplicates.
    \item \textbf{How can we evaluate the de-duplication?}\\
    We propose an evaluation method to evaluate duplicate detection by utilizing the metadata of the Clean MIDI subset of LMD, referred to as LMD-clean in this paper.

    \item \textbf{How many duplicates exist in the dataset?} \\
    Using the duplicate detection method with the best performance, we classify the duplicated files in the LMD. Even with the most conservative threshold of rejection, we find 38,134 duplicated files to be filtered. Finally, we present a filtering list of LMD from both our proposed configuration and the most conservative threshold.\footnote{All training and evaluation code is publicly available: \url{https://github.com/jech2/LMD_Deduplication}}

\end{itemize}


\begin{table}[t]
\centering
\scalebox{0.75}{
\begin{tabular}{@{}c|c|c|c@{}}
\toprule
\textbf{Paper} & \textbf{Dataset Version} & \textbf{Split} &  \textbf{Evaluation} \\
\midrule
MuseGAN\cite{dong2018musegan} & LPD-5-matched & N.A. & .\\
MIDI-Sandwich2\cite{liang2019midi-sandwich2} & LPD-full & N.A. & NLL \\
LakhNES\cite{donahue2019lakhnes} & LMD-full & train, valid & PPL \\
PopMAG\cite{ren2020popmag} & LMD-matched & random & PPL, \musEighth \\
MMM\cite{ens2020mmm} & LMD-full & N.A. & . \\
PiRhDy\cite{liang2020pirhdy} & LMD-full & N.A. & . \\
MMD\cite{ens2021building-metamidi} & LMD-matched & N.A. & . \\
Han et al.\cite{han2022instrument} & LMD-full & N.A. & . \\
\textbf{Museformer}\cite{yu2022museformer} & LMD-full & 8:1:1 & PPL \\
\textbf{MusicBERT}\cite{zeng2021musicbert} & LMD-full & N.A. & . \\
FIGARO\cite{von2022figaro} & LMD-full & 8:1:1 & PPL, \musEighth \\
MIDI2Vec\cite{lisena2022midi2vec} & LMD-matched & 9:1 with CV & . \\
YM2413-MDB\cite{YM2413-MDB} & LMD-full & 9:1:1 & . \\
\textbf{Sulun et al.\cite{sulun2022smg_cve}} & LM(P)D-full, matched & N.A. & . \\
Han et al.\cite{han2023systematic} & LMD-full & N.A. & . \\
Anticipatory\cite{thickstun2023anticipatory} & LMD-full & 87:6:6 & PPL, \musEighth \\
text2midi \cite{bhandariText2midiGeneratingSymbolic2025} & LMD-full (MidiCaps) & N.A. & \musEighth \\

\bottomrule
\end{tabular}}
\caption{Studies that used LMD for training. Studies mentioned the dataset duplication issue are \textbf{bolded}. 
LPD is the piano roll version of LMD suggested by \cite{dong2018musegan}. 
Split strategies are N.A. when not described in the paper. CV means cross-validation. The last column explains whether the authors utilized the dataset during the evaluation. \musEighth\  means that the dataset is used in the listening test.}
\vspace{-3mm}
\label{tab:studies_with_lakh}
\end{table}

\section{Related Works}\label{sec:related_works}
\subsection{LMD and Related Studies}
\label{subsec:lmd and related studies}
LMD \cite{RaffelLMDphd} is a dataset released in 2016 with a method for efficiently matching large-scale MIDI corpus collected from the Internet to the Million Song Dataset (MSD) \cite{thierry2011MSD}. At the time, the dataset was intended to be used for applications such as content-based retrieval, corpus studies of music structure and patterns, and transcription using paired audio and MIDI. 
Diverging from its initially suggested applications, this dataset has become frequently used for training symbolic music generation models, primarily because it is one of the largest available symbolic music datasets. Among the available LMD versions, LMD-full, which contains 178,561 files with unique MD5 hashes, is utilized as the most popular for multi-instrumental pop music generation, and LMD-matched, which consists of 45,129 songs matched with MSD is also used in several studies. 

Since the release of LMD, larger-scale datasets based on web scraping—such as MetaMIDI (MMD)\cite{ens2021building-metamidi} and GigaMIDI \cite{leeGigaMIDIDatasetFeatures2025}—have emerged.  
Notably, the recently released GigaMIDI dataset is a superset of LMD. Additionally, the introduction of the MidiCaps dataset \cite{melechovskyMidiCapsLargescaleMIDI2024}, which incorporates LLM-generated text captions aligned with LMD, has further expanded applications of LMD in text-to-music generation \cite{bhandariText2midiGeneratingSymbolic2025} and music retrieval tasks \cite{wu2024clamp2multimodalmusic, wu2025clamp3universalmusic}.

As shown in Table \ref{tab:studies_with_lakh}, most studies employed the LMD-full and split it for training without mentioning the split strategies. Also, several papers employed the NLL loss or PPL values for their evaluation. We note that a few papers in Table \ref{tab:studies_with_lakh} pointed out the duplication issues within the dataset and removed the duplicated files using a rule-based approach, such as MIDI encoding hash matching. However, we found that the rule-based approach is not sufficient to remove all of the duplicated files in the dataset, which we will show in the evaluation section.
In this study, we used LMD-clean as our de-duplication study and test dataset. This dataset contains 17,184 files organized by the artist and song name in the directory and filename, which we found to have multiple MIDI files of the same song.

\subsection{Dataset De-duplication and Related Issues}
Recently, the need for dataset de-duplication has gained attention across various domains.
In computer vision, \cite{webster2023duplication} proposed a method to eliminate duplication by compressing CLIP features using a contrastive feature compression technique. In music information retrieval (MIR), \cite{batlleroca2024towards} recently introduced a method for detecting exact duplicates in training data using audio-based music similarity metrics. Furthermore, in natural language processing, \cite{lee2021deduplicatingbetter} investigated the effects of dataset de-duplication by applying exact substring matching and hash-based techniques. Their findings showed that removing duplicates improves language model performance, reduces training time, and lowers the rate of training data memorization without harming perplexity. In this work, we focus on the de-duplication method for large-scale symbolic music datasets.

\subsection{Symbolic Music Understanding and Retrieval}
With the advent of large-scale language understanding models such as BERT \cite{devlin2019bert} and BART \cite{lewis2020bart} in the natural language domain, several counterparts have been introduced for symbolic music, including MidiBERT-Piano \cite{midibertpiano}, MusicBERT \cite{zeng2021musicbert}, and PianoBART \cite{pianobart}. Among these, MusicBERT is a large-scale pre-trained model trained on LMD and a private dataset. 
While earlier methods focused solely on the MIDI modality, recent approaches have explored symbolic music retrieval using text, led by the introduction of CLaMP \cite{wu2023clamp}, which learns joint embeddings of symbolic music and text through contrastive learning, using ABC notation as its input format. CLaMP2 \cite{wu2024clamp2multimodalmusic} extends this approach by enhancing the symbolic encoder to support both ABC and MIDI formats with multilingual support. CLaMP3 \cite{wu2025clamp3universalmusic} further generalizes the model to handle additional modalities, including audio and images.
We used these models for our dataset de-duplication task by leveraging their embeddings.

\section{Duplication Types in the LMD}\label{sec:duplication_types_lmd}
We describe the types of duplicated MIDI files observed in LMD-clean. In a broad sense, if we focus on music generation, all arrangements of the same song should be defined as duplication (i.e., the same song cannot be in different splits). However, in tasks such as music arrangement, different arrangements can be considered as different data samples. Considering this aspect, we separated the duplication type into two cases: hard duplication and soft duplication\footnote{We show examples of duplication types in the companion website.}. This work focuses on detecting hard duplication. 

\subsection{Hard Duplication from Similar Arrangement}\label{subsec: hard duplication type}
We define hard duplication as files that share identical sections of arrangements with minor differences. 
These differences include instrument mapping or order, tempo, start offset, file length, missing tracks, or note-level alterations such as pitch, duration, or velocity. Melodic variations may also appear, such as added ornamentation or changes in the number of chord tones played by a specific instrument. Key transpositions are considered hard duplication when other musical elements remain nearly identical but are treated as soft duplication if accompanied by significant stylistic or structural changes.
We assume that some hard duplicates were likely collected as users modified and re-uploaded existing files originally created by other arrangers.

\subsection{Soft Duplication from Different Arrangement}\label{subsec: soft duplication type} 
In soft duplication, MIDI files preserve essential musical elements such as melody, harmony, ornaments, and instrumentation but differ in arrangement style, reflecting the diverse styles of individual arrangers. 
Such duplication includes cases where the core melody remains unchanged or similar, but accompaniment styles (e.g., arpeggio, waltz), pitch ranges, and overall structure vary considerably. In cases of extremely different arrangements, these variations might lead a listener to perceive them as distinct songs.

\section{Duplicate Detection: Dataset and Conventional Approaches}\label{sec:methodology}
Along with the dataset for evaluation, we first explored several approaches for identifying duplicates, including simple rule-based methods and pre-trained symbolic music retrieval models.

\subsection{Dataset}\label{subsec:body}
We used LMD-clean as our evaluation benchmark to assess how well each method detects duplicates within the dataset. LMD-clean is organized by artist folders and song filenames, where duplicate instances of the same song by the same artist are labeled with variations in the filenames (e.g., \texttt{Dancing Queen.mid}, \texttt{Dancing queen.2.mid}). According to this metadata, 10,355 out of 17,184 files in LMD-clean are considered duplicates.

\subsection{Rule-based Approach}
The following rule-based methods serve as baselines for identifying duplicates in the dataset. We assumed that hard duplicated samples share highly similar MIDI-level features. 
Based on this assumption, we explored several methods aimed at detecting and filtering out files with identical or nearly identical features at the beat or pitch level.

\subsubsection{MIDI Encoding Hash}
As discussed in Section \ref{subsec:lmd and related studies}, some studies \cite{zeng2021musicbert, yu2022museformer, sulun2022smg_cve} employed a hash-based approach to detect duplicated MIDI files with different metadata. Here, we used the file de-duplication code of MusicBERT~\cite{zeng2021musicbert}. The string versions of Octuple representations are encoded according to the MD5 hash value, and the hash values of all MIDI files from LMD-clean are compared.

\subsubsection{Beat Position Entropy} 
In our preliminary study, we found there are many duplicates that have exactly the same music but with different instrument mapping or track order. To detect these duplicates, we applied a simple method that checks the distribution of note position within a bar using the MIDI encoding scheme of \cite{ryu2024nested}. 
We computed entropy values from note position distributions at a 16th-note resolution. Files with identical entropy values were identified as hard duplicates.

\subsubsection{Chroma-DTW}
To detect duplicates with similar pitch content, we measured the chroma-level distance between MIDI files. Piano roll-based chromagrams were first generated and aligned by transposing them with the highest pitch occurrence across files. Dynamic Time Warping (DTW) was then applied to measure the similarity between aligned chromagrams. 
To reduce the computational cost of applying DTW to the entire dataset, we first computed pitch histograms for all files and measured the pairwise Kullback-Leibler (KL) divergence. For each file, we selected the top 250 candidates with the lowest KL divergence and then applied chromagram-based DTW to these candidates. Although this approach discards temporal information, it serves as a rough prefiltering step. 

\subsection{Previous Symbolic Music Embedding Models}
We utilized pre-trained MusicBERT \cite{zeng2021musicbert} and CLaMP model series \cite{wu2023clamp, wu2024clamp2multimodalmusic, wu2025clamp3universalmusic} since they support multi-instrumental MIDI. For MusicBERT, we used the pre-trained MusicBERT-small and MusicBERT-base models for inference. For CLaMP, we used the pre-trained CLaMP-512, CLaMP-1024, CLaMP2 and CLaMP3 models. Since the CLaMP-512 and CLaMP-1024 models use XML for input files and ABC notation for their internal data representation, the MIDI files are first converted using Musescore batch processing and then converted with the XML to ABC conversion algorithm. For CLaMP 2 and 3, we converted MIDI to MTF, their MIDI encoding scheme. 

\begin{table}[]
\centering
\scalebox{0.89}{
\begin{tabular}{@{}c|cc@{}}
\toprule
\textbf{Level} & \textbf{Augmentation} & \textbf{Value} \\
\midrule
Note & Onset Shift & (-2, 2) \\
Note & Duration Shift & (-4, 4) \\
Note & Velocity Shift & (-3, 3) \\
Track & Pitch Octave Shift & (-24, 24) \\
Track & Inst Order & Shuffle \\
Track & Inst Mapping & Except Drum \\
Track & Inst Drop & Less than 50\% \\
Track & Bar Drop & 15\% \\
Segment & Bar Shift & (1, 4) \\
Segment & Note Drop & 15\% \\
Segment & Pitch Transpose & (-6, 6) \\
\bottomrule
\end{tabular}}
\caption{List of augmentations for generating MIDI variations. Values in braces are in the token level (inclusive).}
\label{tab:list_of_augmentations}
\vspace{-3mm}
\end{table}

\section{Duplicate Detection: A Contrastive Learning-based Approach}
Previous pre-trained symbolic embedding models were not originally trained for duplicate detection. Inspired by \cite{barnett2024exploring, fradet2023impact} that evaluated the robustness of audio or music embedding models against perturbations such as pitch shift, we explore whether training with such perturbations would improve song identification despite variations. To this end, we developed a BERT-based model and applied various augmentations to the positive samples in contrastive learning, which we refer to as CAugBERT.

\subsection{Data Representation}
We used the LMD-full dataset, excluding all files present in LMD-clean by matching MD5 hash values. The resulting dataset is referred to as LMD-filtered.
We randomly split the LMD-filtered into 98:1:1 ratio and pre-processed the MIDI files with Octuple encoding using MidiTok \cite{miditok2021}. Although we initially held out the remaining 1\% for testing, we did not use it in our final experiments, as we chose to evaluate our model on LMD-clean instead.

\subsection{Data Augmentation}
\label{section:data augmentation}
To reflect the types of variations described in Section \ref{subsec: hard duplication type}, we constructed positive pairs for contrastive learning using different augmentations, referred to as MIDI variation augmentation. Each augmentation of MIDI variation is independently and randomly applied during training. The details of these augmentations are provided in Table \ref{tab:list_of_augmentations}.
In addition, following \cite{han2023systematic}, we used neighbor segments, which are different parts from the same piece, as positive pairs for training. Each piece was first segmented into chunks of 1024 tokens, and then a random segment was selected. MIDI variation augmentation was also applied to neighbor segments to enhance robustness.

\subsection{Model Description}
The implementation of CAugBERT is based on the code from \cite{han2023systematic}, which applies contrastive learning to a BERT architecture. To align with the parameter settings of MusicBERT-small, we used a 4-layer transformer with a sequence length of 1024, hidden size and vocabulary embedding size of 512, and a feedforward dimension of 2048. We use a total batch size of 64 across two A6000 GPUs. 


For masked language modeling (MLM), we adopted the same element, compound, and bar-level masking strategies used in MusicBERT. Contrastive learning was guided by the NT-Xent loss \cite{chen2020simple}. The final loss was computed as a weighted sum of the MLM and contrastive (NT-Xent) losses, with weights of 0.3 and 1.0, respectively.

During training, each encoded MIDI segment was augmented using either MIDI variation or neighbor augmentation described in Section \ref{section:data augmentation}, to maximize the diversity of augmentations within each batch. For validation, fixed manual seed values were used to maintain consistent augmentations across validation batches. To evaluate the effectiveness of the contrastive learning approach for duplication detection, we conducted an ablation study on the contrastive loss, as presented in Table \ref{tab:ranking_metrics}.

\begin{table*}[t]
\centering
\scalebox{0.89 }{
\begin{tabular}{@{}c|cccccc@{}}
\toprule
\textbf{Method} & \textbf{nDCG@all} & \textbf{MRR} & \textbf{Precision} & \textbf{Recall} & \textbf{F1} & \textbf{FN} \\
\midrule
MIDI Encoding Hash      & 0.283 & 0.280 & 0.922 & 0.172 & 0.291 & 6,759 \\
Beat Position Entropy   & 0.344 & 0.329 & 0.919 & 0.192 & 0.318 & 6,435 \\
Chroma-DTW              & 0.157 & 0.081 & 0.826 & 0.050 & 0.094 & 8,730 \\
\midrule
MusicBERT\_small \cite{zeng2021musicbert}       & 0.563 & 0.590 & 0.904 & \underline{0.336} & \underline{0.490} & \underline{4,685} \\
MusicBERT\_base \cite{zeng2021musicbert}        & 0.556 & 0.580 & 0.904 & 0.324 & 0.477 & 4,799 \\
CLaMP-512 \cite{wu2023clamp}               & 0.580 & 0.610 & 0.906 & 0.273 & 0.419 & 5,289 \\
CLaMP-1024 \cite{wu2023clamp}              & \underline{0.646} & \underline{0.673} & 0.903 & 0.317 & 0.469 & 4,762 \\
CLaMP2 \cite{wu2024clamp2multimodalmusic}    & 0.627 & 0.647 & 0.903 & 0.117 & 0.208 & 7,384 \\
CLaMP3 \cite{wu2025clamp3universalmusic}                 & \textbf{0.697} & \textbf{0.709} & 0.902 & 0.210 & 0.341 & 6,017 \\
BERT without Contrastive                    & 0.549 & 0.580 & 0.901 & 0.253 & 0.395 & 5,630 \\
CAugBERT                & 0.558 & 0.586 & 0.903 & \textbf{0.339} & \textbf{0.493} & \textbf{4,623} \\

\bottomrule
\end{tabular}}
 \caption{Overall performances of retrieval and classification metrics on LMD-clean. For nDCG and MRR, we report mean value. 
 For precision, recall, and F1 of neural network-based approaches, the performance with the best threshold (precision > 0.9) is reported. FN represents the false negative, the number of duplicates that the model failed to detect. \textbf{Bold} text indicates the best performance, while \underline{underline} represents the second best.
 }
  \vspace{-3mm}
\label{tab:ranking_metrics}
\end{table*}

\begin{table}[h]
\centering
\renewcommand{\arraystretch}{1.1} 
\scalebox{0.96}{ 
\footnotesize 
\begin{tabular}{@{}ccccc@{}}
\toprule
\textbf{Prediction Strategies} & \textbf{Precision} & \textbf{Recall} & \textbf{F1} & \textbf{FN} \\
\midrule
$\mathcal{M}_{\text{all}}$ 
& 0.881 & 0.411 & \textbf{0.561} &\textbf{3,734} \\
$\mathcal{M}_{\text{rule}}$  & 0.897 & 0.233 & 0.370 & 5,858 \\
MusicBERT $\cup$\; \textnormal{CAugBERT} & 0.901 & 0.369 & 0.523 & 4,300 \\
CLaMP $\cup$\; \textnormal{CAugBERT} * & 0.899 & 0.395 & \underline{0.548} & \underline{3,954} \\
CLaMP $\cup$\; \textnormal{CAugBERT} $\cup$  $\mathcal{M}_{\text{rule}}$
 & 0.888 & 0.395 & 0.547 & 3,917 \\
\bottomrule
\end{tabular}
}
\caption{Classification performance using combinations of available methods on LMD-clean. 
CLaMP: CLaMP-1024 / MusicBERT: MusicBERT\_small / $\mathcal{M}_{\text{all}}$: the union of all methods / $\mathcal{M}_{\text{rule}}$: the subset consisting only of rule-based methods. *: the proposed configuration that we used in the final LMD de-duplication process / FN: the false negative, where duplicates exist but are not predicted / \textbf{Bold}: the best performance / \underline{underline}: the second best.}
\vspace{-10pt}
\label{tab:with_pretrain}
\end{table}

\section{Evaluation}
We evaluated all approaches from two perspectives: (1) Does the system rank duplicates as more similar than others? (2) How accurately does it identify true duplicates? To answer these questions, we utilized the metrics that are commonly used for recommendation and retrieval systems. 


\subsection{Measuring Similarities}
For MIDI Encoding Hash, the similarity between samples was set as 1 when encoding hashes matched. Similarity of beat position entropy was computed by subtracting the absolute difference in entropy from the maximum value of 1. For Chroma-DTW, the similarity was calculated as 1 minus the DTW distance. In the MusicBERT series, similarity was measured using the cosine similarity of the average token embeddings from the Transformer's final hidden layer. For CAugBERT, we used the [CLS] token embedding from the final hidden layer. All BERT-based models utilized 512-dimensional embedding. For CLaMP series, we used a pre-trained 768-dimensional embedding where the last hidden state was average pooled and passed through a projection layer, following the code provided in \cite{wu2023clamp, wu2024clamp2multimodalmusic, wu2025clamp3universalmusic}.

\subsection{Evaluation with Retrieval Metrics}
To evaluate how well each method assigns higher similarity scores to the duplicates, we adopt normalized Discounted Cumulative Gain (nDCG) and Mean Reciprocal Rank (MRR) as evaluation metrics. nDCG measures how highly relevant items are ranked, assigning higher scores when duplicates appear closer to the top of the retrieval list. It is computed by normalizing Discounted Cumulative Gain with the optimal ranking where all duplicates are retrieved at the highest possible ranks. For each query in LMD-clean, we assign relevance 1 to duplicates and 0 to others when computing nDCG. MRR is defined as the average of the inverse ranks of the highest-ranked relevant item for the query. This corresponds to the average rank of the highest similarity samples among the duplicates. 








For the nDCG and MRR metrics, neural network-based approaches outperformed the rule-based methods. Among them, the CLaMP model series consistently achieved higher scores than BERT-based models, with CLAMP3 showing the best overall performance. Since CLaMP models were specifically trained for retrieval tasks, the result is consistent with its intended design. 

However, we observed that all approaches performed below a certain upper bound. In particular, while neural network-based models performed well on hard duplicates, they struggled with detecting soft duplicates. Manual inspection of retrieved samples revealed that similarity scores often failed to reflect perceptual similarity in these soft duplicate cases. For pre-trained symbolic music understanding and retrieval models, we can interpret this as a result of the pre-trained model not being optimized for duplication detection tasks. For the contrastive learning-based approach, detecting soft duplications involving different arrangements appears to be challenging under the current MIDI variation augmentation strategies, as they rely on relatively simple, rule-based augmentations.

\subsection{Evaluation with Classification Metrics}
\label{subsection: evaluation with classification metrics}
We report the precision and recall of each model. While F1-score is commonly used for deciding the best threshold to balance precision and recall, such thresholds may result in an unacceptably high rate of false positives in practical de-duplication. To ensure that our model comparison remains meaningful under realistic scenarios, 
we selected the lowest threshold that satisfies a precision of 0.9.

According to Table \ref{tab:ranking_metrics}, CAugBERT achieved the best performance, which is marginally higher than MusicBERT\_small model. In addition, CAugBERT outperformed the BERT without contrastive loss, which implies that contrastive learning with augmentations is an effective scheme for duplicate detection. In contrast to the retrieval tasks, recent CLaMP models (CLaMP2 and CLaMP3) performed worse than the original CLaMP. We note that BERT-based and CLaMP-based models exhibit different performance trends in both retrieval and classification tasks, implying that the two approaches detect duplication from slightly different perspectives. 

Similar to retrieval metrics, rule-based approaches consistently underperform neural network-based models. In particular, Chroma-DTW failed to achieve high precision across all thresholds. We also note that the precision of MIDI Encoding Hash is not 1.0, which implies that several files in LMD-clean incorrectly matched the song name label; we found a few songs with the exact same MIDI content were matched with different song names. Upon manually reviewing the corresponding recordings, we identified two distinct issues: 1) the metadata of some MIDI files was incorrectly linked to entirely different pieces, and 2) a song appeared multiple times with different metadata, often due to variations in song titles across international releases.

\subsection{Proposed Method for De-duplication}
\label{subsec:proposed method}
Prior to performing the de-duplication of LMD, we investigated whether combining multiple methods could further enhance performance. The results of these experiments are presented in Table \ref{tab:with_pretrain}. Given the differing perspectives of each method, we hypothesized that they may detect duplication based on complementary criteria. To test this, we evaluated all possible method combinations with CAugBERT. Among all two-model combinations, the union of CAugBERT and CLaMP-1024 yielded the highest F1 score. While MusicBERT\_small achieved the second-highest F1 score in Table \ref{tab:ranking_metrics}, its union with CAugBERT resulted in a lower F1 score than the union with CLaMP. 
Notably, this combination achieved performance comparable to that of combining all available methods, suggesting that these two models cover the majority of duplicate predictions. While we also explored incorporating rule-based methods into the final ensemble, the performance gain was negligible. Consequently, we selected the union of the CLaMP-1024 and CAugBERT as our proposed configuration for LMD de-duplication.

\begin{table}[t]
\centering
\scalebox{0.95}{ 

\begin{tabular}{c|cc}
\toprule
\textbf{Methods} & \textbf{\# Clusters} & \textbf{\# Duplicates} \\
\midrule
MIDI Encoding Hash \cite{zeng2021musicbert} & 16\,633 & 26\,167 \\
\midrule
Proposed Configuration & 23\,566 & 68\,075 \\
Conservative Configuration & 20\,797 & 38\,134 \\
\bottomrule
\end{tabular}
}
 \caption{Duplicate count in LMD-full when queried with LMD-full, using two configurations. Proposed refers to the configuration described in Section \ref{subsec:proposed method}. Conservative refers to the configuration using embedding similarity ≥0.99. 
 }
 \vspace{-3mm}
\label{tab: number of duplicates}
\end{table}

\begin{figure}[t]
 \centerline{
 \includegraphics[width=\columnwidth]{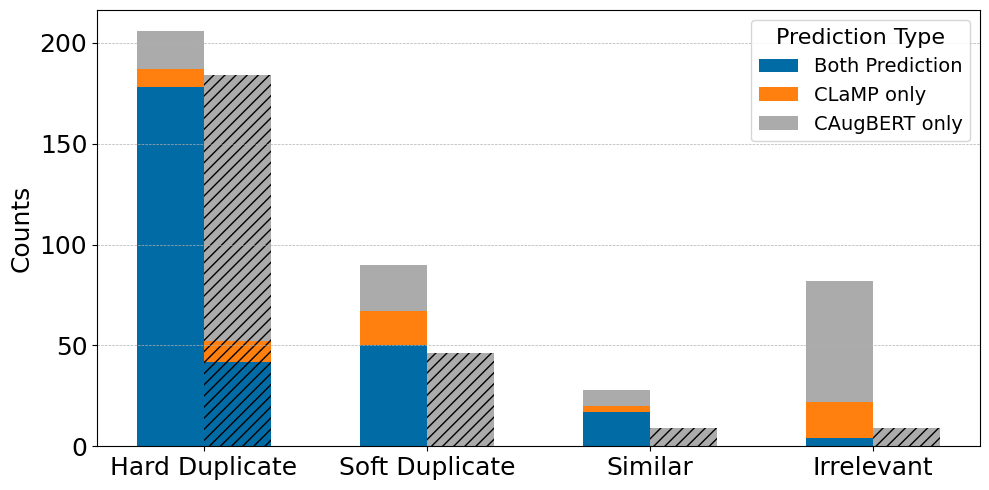}}
 \vspace{-2mm}
 \caption{Listening test on 100 random LMD-full samples with detected duplicates. X-axis shows classified categories from Section \ref{section: deduplication}. Unhatched bars use the proposed threshold; hatched bars use the stricter ≥0.99 threshold.}
 \vspace{-3mm}
\label{fig:listening test}
\end{figure}

To assess how well these thresholds determined on LMD-clean generalize to LMD-full, we conducted a manual validation of the retrieval results. We first extracted all embeddings of LMD-full using the proposed configuration and computed the similarity between embeddings. Duplicate pairs were identified using the thresholds applied in the previous experiment.
We then randomly sampled 100 MIDI files, each containing at least one detected duplicate, resulting in a total of 506 MIDI samples including the 100 queries. After, one of the authors with a bachelor's degree in music composition manually listened to the retrieved items and classified the retrieved samples into four categories. As described in Figure \ref{fig:listening test}, we found that 72.9\% of items are duplicates (soft and hard), whereas almost similar in instrument combinations and chord progressions but not duplicate items were 6.9\%, and the percentage of irrelevant items is 20.2\%. Upon closer inspection, 79.3\% of the irrelevant items turned out to be either fugue-like pieces or short single-instrument segments—both types of data that were underrepresented during evaluating on LMD-clean. When increasing the threshold up to 0.99, the proportion of similar and irrelevant items dropped to 2.22\% and 2.22\%. 

\section{De-duplication of LMD}
\label{section: deduplication}
After we thresholded the similarity, we constructed an adjacency list, whereas detected duplication is an edge, and each file can be a node in a graph. Afterward, we ran a depth-first search to find the clusters of duplication (i.e., connected components of the graph) and generated a duplicated file list except for one sample with the highest total note counts in each cluster. After the clustering process, we found that the number of clusters with duplicates is 23,566, and the number of duplicate files is 68,075, which is the number of files to be filtered, as shown in Table \ref{tab: number of duplicates}. 

Given the performance limitation of the current configuration, we offer three de-duplication options for LMD as our best available solution. 

\begin{figure}[t]
 \centerline{
 \includegraphics[width=\columnwidth]{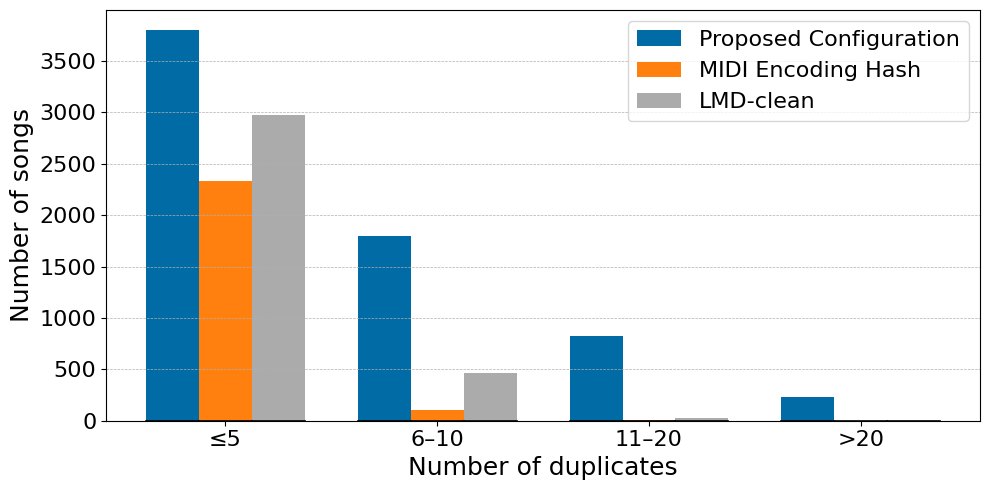}}
 \vspace{-2mm}
 \caption{Duplicate count in LMD-full when queried with LMD-clean, using three different methods. LMD-clean duplicates based on the artist folder and filename. }
 \vspace{-10pt}
\label{fig:dup_song_cnt_in_LMD_clean}
\end{figure}

The first option is filtering pieces based only on the LMD-clean, as the primary use case for LMD is generating multi-instrumental pop music. LMD-clean covers various types of famous songs, which takes a large portion of duplicates in LMD.
We release a duplicate list generated by querying each piece of LMD-clean against LMD-full using our best configuration.  The number of detected duplicates is shown in Figure \ref{fig:dup_song_cnt_in_LMD_clean}. This will help researchers effectively remove highly duplicated popular tracks. 

Also, to address diverse use cases, we provide duplicate lists by querying from the entire LMD-full, using both the proposed configuration and a more conservative threshold (0.99 for both models). These options are designed to support scenarios that either prioritize aggressive duplicate removal or aim to minimize the elimination of irrelevant items. We note that even with a conservative threshold of rejection of 0.99, it yielded 20,797 duplicate clusters with 38\,134 duplicated files to be filtered, which is larger than MusicBERT's file encoding hash detects: 26,167.

\section{Conclusion}
In this study, we highlighted dataset duplication issues in MIR and examined the capacities of various approaches for de-duplicating the LMD. As a result, we found that the 38\,134 files in the LMD-full (21.4\%) are considered as duplications with high confidence with the proposed configuration. The resulting de-duplication lists and our approaches can enhance the overall validity of symbolic music research. We expect the methods we explored can be applied to other large-scale symbolic music datasets.

Future work includes investigating the impact of de-duplication on the training and evaluation of MIR models, similar to the analysis of training data attribution in audio-based music generation by \cite{barnett2024exploring}. Improving soft duplicate detection while minimizing irrelevant matches also remains an open challenge. One promising direction is to leverage neural methods for MIDI variation generation \cite{han2022instrument, huertavarynote} to construct positive pairs for contrastive learning.


\section{Acknowledgements}
This work was supported by the National Research Foundation of Korea (NRF) grant funded by the Korea government (MSIT) (RS-2025-00560548).
\bibliography{ISMIR2025_template}

\begin{thebibliography}{10}
\providecommand{\url}[1]{#1}
\csname url@samestyle\endcsname
\providecommand{\newblock}{\relax}
\providecommand{\bibinfo}[2]{#2}
\providecommand{\BIBentrySTDinterwordspacing}{\spaceskip=0pt\relax}
\providecommand{\BIBentryALTinterwordstretchfactor}{4}
\providecommand{\BIBentryALTinterwordspacing}{\spaceskip=\fontdimen2\font plus
\BIBentryALTinterwordstretchfactor\fontdimen3\font minus \fontdimen4\font\relax}
\providecommand{\BIBforeignlanguage}[2]{{%
\expandafter\ifx\csname l@#1\endcsname\relax
\typeout{** WARNING: IEEEtran.bst: No hyphenation pattern has been}%
\typeout{** loaded for the language `#1'. Using the pattern for}%
\typeout{** the default language instead.}%
\else
\language=\csname l@#1\endcsname
\fi
#2}}
\providecommand{\BIBdecl}{\relax}
\BIBdecl

\bibitem{jia2021scaling}
C.~Jia, Y.~Yang, Y.~Xia, Y.-T. Chen, Z.~Parekh, H.~Pham, Q.~Le, Y.-H. Sung, Z.~Li, and T.~Duerig, ``Scaling up visual and vision-language representation learning with noisy text supervision,'' in \emph{Proceedings of the 38th International Conference on Machine Learning (ICML)}, 2021.

\bibitem{RaffelLMDphd}
C.~Raffel, ``Learning-based methods for comparing sequences, with applications to audio-to-midi alignment and matching,'' Ph.D. dissertation, Columbia University, 2016.

\bibitem{manilow2019cutting_slakh}
E.~Manilow, G.~Wichern, P.~Seetharaman, and J.~Le~Roux, ``Cutting music source separation some {Slakh}: A dataset to study the impact of training data quality and quantity,'' in \emph{Proceedings of IEEE Workshop on Applications of Signal Processing to Audio and Acoustics (WASPAA)}, 2019.

\bibitem{webster2023duplication}
R.~Webster, J.~Rabin, L.~Simon, and F.~Jurie, ``On the de-duplication of {LAION}-2b,'' \emph{arXiv preprint: 2303.12733}, 2023.

\bibitem{lee2021deduplicatingbetter}
K.~Lee, D.~Ippolito, A.~Nystrom, C.~Zhang, D.~Eck, C.~Callison-Burch, and N.~Carlini, ``Deduplicating training data makes language models better,'' in \emph{Proceedings of the 60th Annual Meeting of the Association for Computational Linguistics (Volume 1: Long Papers)}, 2022.

\bibitem{sturm2013gtzan}
B.~L. Sturm, ``The gtzan dataset: Its contents, its faults, their effects on evaluation, and its future use,'' \emph{arXiv preprint:1306.1461}, 2013.

\bibitem{sturm2014state}
------, ``The state of the art ten years after a state of the art: Future research in music information retrieval,'' \emph{Journal of new music research}, vol.~43, no.~2, pp. 147--172, 2014.

\bibitem{liang2019midi-sandwich2}
X.~Liang, J.~Wu, and J.~Cao, ``{MIDI-Sandwich2}: Rnn-based hierarchical multi-modal fusion generation vae networks for multi-track symbolic music generation,'' \emph{arXiv preprint: 1909.03522}, 2019.

\bibitem{donahue2019lakhnes}
C.~Donahue, H.~H. Mao, Y.~E. Li, G.~W. Cottrell, and J.~McAuley, ``{LakhNES}: Improving multi-instrumental music generation with cross-domain pre-training,'' in \emph{Proceedings of the 20th International Society for Music Information Retrieval Conference}, 2019.

\bibitem{ren2020popmag}
Y.~Ren, J.~He, X.~Tan, T.~Qin, Z.~Zhao, and T.-Y. Liu, ``{PopMAG}: Pop music accompaniment generation,'' in \emph{Proceedings of the 28th ACM International Conference on Multimedia}, 2020.

\bibitem{yu2022museformer}
B.~Yu, P.~Lu, R.~Wang, W.~Hu, X.~Tan, W.~Ye, S.~Zhang, T.~Qin, and T.-Y. Liu, ``Museformer: Transformer with fine- and coarse-grained attention for music generation,'' in \emph{Advances in Neural Information Processing Systems 35 (NeurIPS 2022)}, 2022.

\bibitem{von2022figaro}
D.~von R{\"u}tte, L.~Biggio, Y.~Kilcher, and T.~Hofmann, ``Figaro: Generating symbolic music with fine-grained artistic control,'' \emph{Proceedings of the International Conference on Learning Representations (ICLR)}, 2023.

\bibitem{thickstun2023anticipatory}
J.~Thickstun, D.~L.~W. Hall, C.~Donahue, and P.~Liang, ``Anticipatory {Music} {Transformer},'' \emph{Transactions on Machine Learning Research}, 2024.

\bibitem{bhandariText2midiGeneratingSymbolic2025}
K.~Bhandari, A.~Roy, K.~Wang, G.~Puri, S.~Colton, and D.~Herremans, ``Text2midi: Generating symbolic music from captions,'' in \emph{Proceedings of the 39th AAAI Conference on Artificial Intelligence}, 2025.

\bibitem{dong2018musegan}
H.-W. Dong, W.-Y. Hsiao, L.-C. Yang, and Y.-H. Yang, ``{Musegan}: Multi-track sequential generative adversarial networks for symbolic music generation and accompaniment,'' in \emph{Proceedings of the 32th AAAI Conference on Artificial Intelligence}, vol.~32, no.~1, 2018.

\bibitem{ens2020mmm}
J.~Ens and P.~Pasquier, ``{MMM}: Exploring conditional multi-track music generation with the transformer,'' \emph{arXiv preprint: 2008.06048}, 2020.

\bibitem{liang2020pirhdy}
H.~Liang, W.~Lei, P.~Y. Chan, Z.~Yang, M.~Sun, and T.-S. Chua, ``{Pirhdy}: Learning pitch-, rhythm-, and dynamics-aware embeddings for symbolic music,'' in \emph{Proceedings of the 28th ACM International Conference on Multimedia}, 2020.

\bibitem{ens2021building-metamidi}
J.~Ens and P.~Pasquier, ``Building the {MetaMIDI} dataset: Linking symbolic and audio musical data.'' in \emph{Proceedings of the 22nd International Society for Music Information Retrieval Conference}, 2021.

\bibitem{han2022instrument}
S.~Han, H.~Ihm, D.~Ahn, and W.~Lim, ``Instrument separation of symbolic music by explicitly guided diffusion model,'' \emph{Proceedings of the NeurIPS Workshop on Machine Learning for Creativity and Design}, 2022.

\bibitem{zeng2021musicbert}
M.~Zeng, X.~Tan, R.~Wang, Z.~Ju, T.~Qin, and T.-Y. Liu, ``{MusicBERT}: Symbolic music understanding with large-scale pre-training,'' in \emph{Findings of the Association for Computational Linguistics: ACL-IJCNLP 2021}, 2021.

\bibitem{lisena2022midi2vec}
P.~Lisena, A.~Mero{\~n}o-Pe{\~n}uela, and R.~Troncy, ``Midi2vec: Learning midi embeddings for reliable prediction of symbolic music metadata,'' \emph{Semantic Web}, vol.~13, no.~3, pp. 357--377, 2022.

\bibitem{YM2413-MDB}
E.~Choi, Y.~Chung, S.~Lee, J.~Jeon, T.~Kwon, and J.~Nam, ``{YM2413-MDB}: A multi-instrumental {FM} video game music dataset with emotion annotations,'' in \emph{Proceedings of the 23rd International Society for Music Information Retrieval Conference}, 2022.

\bibitem{sulun2022smg_cve}
S.~Sulun, M.~E. Davies, and P.~Viana, ``Symbolic music generation conditioned on continuous-valued emotions,'' \emph{IEEE Access}, vol.~10, pp. 44\,617--44\,626, 2022.

\bibitem{han2023systematic}
S.~Han, H.~Ihm, and W.~Lim, ``Systematic analysis of music representations from bert,'' \emph{arXiv preprint arXiv:2306.04628}, 2023.

\bibitem{thierry2011MSD}
T.~Bertin-Mahieux, D.~P.~W. Ellis, B.~Whitman, and P.~Lamere, ``{The Million Song Dataset},'' in \emph{Proceedings of the 12th International Society for Music Information Retrieval Conference}, 2011.

\bibitem{leeGigaMIDIDatasetFeatures2025}
K.~J.~M. Lee, J.~Ens, S.~Adkins, P.~Sarmento, M.~Barthet, and P.~Pasquier, ``The {GigaMIDI} dataset with features for expressive music performance detection,'' \emph{Transactions of the International Society for Music Information Retrieval}, vol.~8, no.~1, pp. 1--19, 2025.

\bibitem{melechovskyMidiCapsLargescaleMIDI2024}
J.~Melechovsky, A.~Roy, and D.~Herremans, ``{{MidiCaps}}: A large-scale midi dataset with text captions,'' in \emph{Proceedings of the 25th International Society for Music Information Retrieval Conference}, 2024.

\bibitem{wu2024clamp2multimodalmusic}
S.~Wu, Y.~Wang, R.~Yuan, Z.~Guo, X.~Tan, G.~Zhang, M.~Zhou, J.~Chen, X.~Mu, Y.~Gao, Y.~Dong, J.~Liu, X.~Li, F.~Yu, and M.~Sun, ``{CLaMP 2}: Multimodal music information retrieval across 101 languages using large language models,'' \emph{arXiv preprint: 2410.13267}, 2024.

\bibitem{wu2025clamp3universalmusic}
S.~Wu, Z.~Guo, R.~Yuan, J.~Jiang, S.~Doh, G.~Xia, J.~Nam, X.~Li, F.~Yu, and M.~Sun, ``{CLaMP 3}: Universal music information retrieval across unaligned modalities and unseen languages,'' \emph{arXiv preprint: 2502.10362}, 2025.

\bibitem{batlleroca2024towards}
R.~Batlle-Roca, W.-H. Liao, X.~Serra, Y.~Mitsufuji, and E.~Gómez, ``Towards assessing data replication in music generation with music similarity metrics on raw audio,'' in \emph{Proceedings of the 25th International Society for Music Information Retrieval Conference}, 2024.

\bibitem{devlin2019bert}
J.~Devlin, M.-W. Chang, K.~Lee, and K.~Toutanova, ``{BERT: Pre-training of Deep Bidirectional Transformers for Language Understanding},'' in \emph{Proceedings of the 2019 Conference of the North American Chapter of the Association for Computational Linguistics: Human Language Technologies, Volume 1 (Long and Short Papers)}, 2019.

\bibitem{lewis2020bart}
M.~Lewis, Y.~Liu, N.~Goyal, M.~Ghazvininejad, A.~Mohamed, O.~Levy, V.~Stoyanov, and L.~Zettlemoyer, ``{BART}: Denoising sequence-to-sequence pre-training for natural language generation, translation, and comprehension,'' in \emph{Proceedings of the 58th Annual Meeting of the Association for Computational Linguistics}, 2020.

\bibitem{midibertpiano}
Y.-H. Chou, I.-C. Chen, J.~Ching, C.-J. Chang, and Y.-H. Yang, ``Midibert-piano: Large-scale pre-training for symbolic music classification tasks,'' \emph{Journal of Creative Music Systems}, vol.~8, no.~1, 2024.

\bibitem{pianobart}
X.~Liang, Z.~Zhao, W.~Zeng, Y.~He, F.~He, Y.~Wang, and C.~Gao, ``Pianobart: Symbolic piano music generation and understanding with large-scale pre-training,'' in \emph{Proceedings of the 2024 IEEE International Conference on Multimedia and Expo (ICME)}, 2024.

\bibitem{wu2023clamp}
S.~Wu, D.~Yu, X.~Tan, and M.~Sun, ``{CLaMP: Contrastive Language-Music Pre-training for Cross-Modal Symbolic Music Information Retrieval},'' in \emph{Proceedings of the 24th International Society for Music Information Retrieval Conference}, 2023.

\bibitem{ryu2024nested}
J.~Ryu, H.-W. Dong, J.~Jung, and D.~Jeong, ``Nested music transformer: Sequentially decoding compound tokens in symbolic music and audio generation,'' in \emph{Proceedings of the 25th International Society for Music Information Retrieval Conference}, 2024.

\bibitem{barnett2024exploring}
J.~Barnett, H.~F. Garcia, and B.~Pardo, ``Exploring musical roots: Applying audio embeddings to empower influence attribution for a generative music model,'' in \emph{Proceedings of the 25th International Society for Music Information Retrieval Conference}, 2024.

\bibitem{fradet2023impact}
N.~Fradet, N.~Gutowski, F.~Chhel, and J.-P. Briot, ``Impact of time and note duration tokenizations on deep learning symbolic music modeling,'' in \emph{Proceedings of the 24th International Society for Music Information Retrieval Conference}, 2023.

\bibitem{miditok2021}
N.~Fradet, J.-P. Briot, F.~Chhel, A.~El~Fallah~Seghrouchni, and N.~Gutowski, ``{MidiTok}: A python package for {MIDI} file tokenization,'' in \emph{Extended Abstracts for the Late-Breaking Demo Session of the 22nd International Society for Music Information Retrieval Conference}, 2021.

\bibitem{chen2020simple}
T.~Chen, S.~Kornblith, M.~Norouzi, and G.~Hinton, ``A simple framework for contrastive learning of visual representations,'' in \emph{Proceedings of the 37th International Conference on Machine Learning (ICML)}, 2020.

\bibitem{huertavarynote}
J.~Huerta, B.~Liu, and P.~Stone, ``Varynote: A method to automatically vary the number of notes in symbolic music,'' in \emph{Bridge after the turmoil - The 16th International Symposium, CMMR 2023}, 2023.

\end{thebibliography}

%
%
%
%

\end{document}